\documentclass[twocolumn,showpacs,preprintnumbers,showkeys,superscriptaddress]{revtex4}
\usepackage{amssymb}
\usepackage{graphicx}
\usepackage{dcolumn}
\usepackage{bm}
\usepackage{appendix}

\def\eqref#1{Eq.~(\ref{eq:#1})}

\begin{document}

\title{A Number-Conserving Theory for Nuclear Pairing}
\author{L. Y. Jia}  
\affiliation{Department of Physics, University of Shanghai for
Science and Technology, Shanghai 200093, P. R. China}
\affiliation{Department of Physics, Hebei Normal University,
Shijiazhuang, Hebei 050024, P. R. China}

\date{\today}

\begin{abstract}

A microscopic theory for nuclear pairing is proposed through the
generalized density matrix formalism. The analytical equations are
as simple as that of the BCS theory, and could be solved within a
similar computer time. The current theory conserves the exact
particle number, and is valid at arbitrary pairing strength
(including those below the BCS critical strength). These are the two
main advantages over the conventional BCS theory. The theory is also
of interests to other mesoscopic systems.

\end{abstract}


\vspace{0.4in}

\maketitle

\section{Introduction}

The BCS theory is first proposed as a microscopic theory for
superconductivity \cite{BCS}. Later it is adopted in nuclear physics
for treating pairing correlations \cite{BCS_nucl1, BCS_nucl2}. After
fifty years, it is still the ``standard'' treatment (see Ref.
\cite{BCS_book}), mainly because of its simplicity and the
convenience in adding higher-order correlations (for example by
QRPA). However, there are two main disadvantages of the theory
applied to the finite nucleus, as compared to macroscopic quantum
systems. Firstly, by introducing quasi-particles, it destroys
particle number conservation. Quite often, the fluctuation in
particle number was not small relative to its average value.
Secondly, for the nuclear system with finite level spacing, the BCS
theory requires a minimal pairing strength. Below that strength it
gives only trivial (vanishing) solutions, while in reality the
pairing always has an effect.

The current treatment by the generalized density matrix (GDM) method
does not have the above two deficiencies. Yet it is simple enough
for further treatment of higher-order correlations within the same
GDM framework. We will first present the formalism in Sec.
\ref{sec_Form}. Next the theory is applied to calcium isotopes in
Sec. \ref{Sec_Ca} with comparisons to the exact shell-model results
and that of BCS. At last Sec. \ref{Sec_sum} summarizes the work and
discusses further directions.

\section{Formalism  \label{sec_Form}}

The GDM formalism was originally introduced in Refs.
\cite{kerman63,BZ,Zele_ptps,shtokman75} and recently reconsidered in
Refs. \cite{Jia_1,Jia_2,Jia_3}. Until now its treatment of nuclear
pairing correlations is limited to the conventional BCS, thus has
the above discussed disadvantages. Here we explore the possibility
of using the ``pair condensate'' (\ref{gs}) (with definite particle
number) as the ``variational'' ground state, instead of the BCS
``quasi-particle vacuum''. Below we set up the GDM formalism in a
general way, but solve in this work only the lowest-order
(mean-field) equations.

We assume that the ground state of the $2N$-particle system is a
$N$-pair condensate,
\begin{eqnarray}
|\phi_N\rangle = \frac{1}{\sqrt{\chi_{N}}} (P^\dagger)^{N} |0\rangle
, \label{gs}
\end{eqnarray}
where $\chi_{N}$ is a normalization factor that will be specified
later [see Eq. (\ref{norm})], and $P^\dagger$ is the pair creation
operator
\begin{eqnarray}
P^\dagger = \frac{1}{2} \sum_1 v_1 a_1^\dagger a_{\tilde{1}}^\dagger
. \label{P_dag}
\end{eqnarray}
In Eq. (\ref{P_dag}) the summation runs over the entire
single-particle space. The pair structure $v_1$ are parameters to be
determined by the theory.\\

With the antisymmetrized fermionic Hamiltonian
\begin{eqnarray}
H = \sum_{12} \epsilon_{12} a_1^\dagger a_2 + \frac{1}{4}
\sum_{1234} V_{1234} a_1^\dagger a_2^\dagger a_3 a_4 ,  \label{H_f}
\end{eqnarray}
we calculate the equations of motion for the one-body density matrix
operators, $R_{12} \equiv a_2^\dagger a_1$ and $K_{12} = a_2 a_1$,
\begin{eqnarray}
[R_{12}, H] = [f\{R\}, R]_{12} - (K \Delta^\dagger\{K\})_{12} +
(\Delta\{K\} K^\dagger)_{12} ,     \label{eom_R} \\
{[} K_{12} , H ] = \Delta\{K\}_{12} + (K f^T\{R\})_{12} + (f\{R\} K)_{12}  \nonumber \\
- (\Delta\{K\} R^T)_{12} - (R \Delta\{K\})_{12} , \label{eom_K}
\end{eqnarray}
where the self-consistent fields are defined as
\begin{eqnarray}
W\{R\}_{12} = \sum_{34} V_{1432} R_{34} ,~~~ f\{R\} = \epsilon + W\{R\} ,  \label{W_field} \\
\Delta\{K\}_{12} = \frac{1}{2} \sum_{34} V_{1234} K_{43}
\label{D_field} .
\end{eqnarray}
On the right-hand side of Eqs. (\ref{eom_R}) and (\ref{eom_K}) we
have used the factorization
\begin{eqnarray}
a_4^\dagger a_3^\dagger a_2 a_1 \circeq a_4^\dagger a_1 \cdot
a_3^\dagger a_2 - a_4^\dagger a_2 \cdot a_3^\dagger a_1 +
a_4^\dagger a_3^\dagger \cdot a_2 a_1 ,   \label{fac1}  \\
a_4^\dagger a_3 a_2 a_1 \circeq a_4^\dagger a_1 \cdot a_3 a_2 -
a_4^\dagger a_2 \cdot a_3 a_1 + a_4^\dagger a_3 \cdot a_2 a_1 ,
\label{fac2}
\end{eqnarray}
generalizing Eq. (11) in Ref. \cite{Jia_2}. In the presence of the
pair condensate terms like $a_4^\dagger a_3^\dagger \cdot a_2 a_1$
are not small. As before ``$\circeq$'' is used when an equation
holds in the collective subspace but not in the full many-body
space.

The method assumes that the Hamiltonian and the density matrix
operators can be expanded as Taylor series of the bosonic mode
operators (collective coordinate $\alpha$ and momentum $\pi$) within
the collective subspace,
\begin{eqnarray}
H \circeq \sum_{ml} \Lambda^{(m,2l)} \frac{1}{2} \frac{\{ \alpha^m ,
\pi^{2l} \}}{m! (2l)!} ,     \label{H_exp}
\end{eqnarray}
and
\begin{eqnarray}
R_{12} = a_2^\dagger a_1 \circeq \sum_{mn} r^{(mn)}_{12} \frac{1}{2}
\frac{\{ \alpha^m , \pi^n
\}}{m! n!} ,     \label{R_exp}  \\
K_{12} = a_2 a_1 \circeq \sum_{mn} k^{(mn)}_{12} \frac{1}{2}
\frac{\{ \alpha^m , \pi^n \}}{m! n!} .     \label{K_exp}
\end{eqnarray}
In Eq. (\ref{K_exp}) $K_{12}$ destroys two particles, hence it
connects the collective subspace with $2N$ particles to that with
$2N-2$ particles. The first term $k^{(00)} \equiv \kappa$ is the
usual ``pair transition amplitude'' between the ground states of
neighboring even-even nuclei. Higher-order terms $k^{(mn)}$
represent the transition amplitudes between the collectively excited
states (with phonons). Strictly speaking, the generalized density
matrices ($r_{N,12}$, $k_{N,12}$), the mode operators ($\alpha_N$,
$\pi_N$), and the bosonic Hamiltonian parameters $\Lambda^{(m,2l)}$
should have the label of particle number $2 N$, and the GDM
equations should be solved simultaneously for all the nuclei between
two magic numbers, in a way similar to that in Ref.
\cite{Zele_pairing}. However in this work we will drop the label
$N$, assuming neighboring even-even nuclei have similar collective
modes ($\alpha_{N} \approx \alpha_{N-1}$, $\pi_{N} \approx
\pi_{N-1}$) and density matrices ($r_{N} \approx r_{N-1}$, $k_{N}
\approx k_{N-1}$). More careful treatment with explicit label $N$
will be discussed in the future.

Substituting the expansions (\ref{H_exp}), (\ref{R_exp}) and
(\ref{K_exp}) into the equations of motion (\ref{eom_R}) and
(\ref{eom_K}), calculating commutators of bosonic operators $\alpha$
and $\pi$, we arrive at the GDM set of equations. In this work we
consider only the lowest-order (mean-field) equations:
\begin{eqnarray}
0 = [f, \rho] - \kappa \delta^\dagger + \delta \kappa^\dagger ,  \label{eom_R0}     \\
(\Lambda^{(00)}_N - \Lambda^{(00)}_{N-1}) \kappa = f \kappa + \delta
- \delta \rho^T - \rho \delta + \kappa f^T , \label{eom_K0}
\end{eqnarray}
where $\rho \equiv r^{(00)}$, $\kappa \equiv k^{(00)}$, $f =
\epsilon + W\{\rho\}$, and $\delta = \Delta\{\kappa\}$ are leading
terms in the expansions of respective quantities
(\ref{W_field},\ref{D_field},\ref{R_exp},\ref{K_exp}).
$\Lambda^{(00)}_N$, the leading term in the bosonic Hamiltonian
(\ref{H_exp}), is the binding energy of the $N$-pair condensate
(\ref{gs}). Usually the difference $\Lambda^{(00)}_N -
\Lambda^{(00)}_{N-1}$ is not small and should be kept. \\

On the ground state (\ref{gs}), $\rho$ and $\kappa$ are
``diagonal'':
\begin{eqnarray}
\rho_{12} = \langle \phi_{N} | a_2^\dagger a_1 | \phi_{N} \rangle  =
\delta_{12} n_1 ,  \label{rho_diag} \\
\kappa_{12} = \langle \phi_{N-1} | a_2 a_1 | \phi_{N} \rangle =
\delta_{\tilde{1}2} s_1 , \label{kappa_diag}
\end{eqnarray}
where $s_1$ and $n_1$ are functions of the pair structure $v$
(\ref{P_dag}), given later by the recursive formula (\ref{ns_eq}).
In a realistic shell-model calculation, usually each single-particle
level has distinct spin and parity, thus both $f$ and $\delta$ are
``diagonal'':
\begin{eqnarray}
f_{12} = \delta_{12} e_1 ,  \label{f_diag} \\
\delta_{12} = \delta_{1\tilde{2}} g_1 .     \label{delta_diag}
\end{eqnarray}
Under Eqs.
(\ref{rho_diag},\ref{kappa_diag},\ref{f_diag},\ref{delta_diag}), Eq.
(\ref{eom_R0}) is satisfied automatically, and Eq. (\ref{eom_K0})
becomes
\begin{eqnarray}
\Lambda^{(00)}_N - \Lambda^{(00)}_{N-1} = 2 e_1 + g_1 \frac{2 n_1 -
1}{s_1} .  \label{eq_final}
\end{eqnarray}
Equation (\ref{eq_final}) is the main equation of the theory. It
implies that the right-hand side is independent of the
single-particle label $1$, which gives $\Omega - 1$ constraints for
a single-particle space of dimension $2 \Omega$ ($\Omega$
time-reversal pairs). These constraints fix the $\Omega - 1$
parameters in Eq. (\ref{P_dag}) (a common factor in $v_1$ does not
matter), which completes the theory. Notice that Eq.
(\ref{eq_final}) has non-trivial (``non-zero'') solution at
infinitesimal pairing (infinitesimal $g_1$).\\

At last we supply the formula for the recursive calculation of
$\rho$ (\ref{rho_diag}) and $\kappa$ (\ref{kappa_diag}) in terms of
$v$ (\ref{P_dag}). Introducing $P_1^\dagger = a_1^\dagger
a_{\tilde{1}}^\dagger$ and
\begin{eqnarray}
t^1_{N} = \langle 0 | P^{N-1} P_1 (P^\dagger)^{N} | 0 \rangle ,
\end{eqnarray}
it is easy to deduce the recursive formula
\begin{eqnarray}
t^1_{N} = \frac{1}{2} N \cdot v_1 \sum_2 v_2 t^2_{N-1} - N (N-1)
\cdot (v_1)^2 t^1_{N-1} ,     \label{t_recur}
\end{eqnarray}
with initial value $t^1_{N=1} = v_1$. Then the normalization factor
$\chi_N$ in Eq. (\ref{gs}) is expressed in terms of $t_N$ as
\begin{eqnarray}
\chi_N = \frac{1}{2} \sum_1 v_1 t^1_{N} .  \label{norm}
\end{eqnarray}
$t_N^1$ and $\chi_N$ are polynomials of $v$. Finally the expressions
for $n_1$ (\ref{rho_diag}) and $s_1$ (\ref{kappa_diag}) are
\begin{eqnarray}
n_1 = \frac{ N v_1 t^1_{N} }{\chi_{N}} ,~~~  s_1 =
\frac{t^1_{N}}{\sqrt{\chi_{N} \chi_{N-1}}} .  \label{ns_eq}
\end{eqnarray}
The functional forms of $n$ and $s$ in terms of $v$ (\ref{P_dag})
are ``kinematics'' of the system (like the ``kinematic''
Clebsch-Gordan coefficients for rotational symmetry), which can be
calculated (and stored or tabulated) once for all for a given model
space. The main computing-time cost of the method should be that to
solve Eq. (\ref{eq_final}). In fact, Eq. (\ref{eq_final}) is a
better behaved equation compared to the BCS equation. It involves
essentially ratio of two polynomials but no square roots.

As a simple check we consider the degenerate pairing model.
Equations (\ref{t_recur}) and (\ref{norm}) become $t_N = N v^2
(\Omega-N+1) t_{N-1}$ and $\chi_N = \Omega v t_{N}$, which in turn
gives $n = N/\Omega$ and $s = \sqrt{N(\Omega-N+1)}/\Omega$ according
to Eq. (\ref{ns_eq}). They agree with the known results. The
right-hand side of Eq. (\ref{eq_final}) becomes $2\epsilon + (G
\Omega s) (\frac{2 N/\Omega - 1}{s}) = 2\epsilon + G (2 N -
\Omega)$, which is the correct binding energy difference
$\Lambda^{(00)}_N - \Lambda^{(00)}_{N-1}$.

\section{Realistic Applications     \label{Sec_Ca}}

We apply the theory to calcium isotopes, using the well established
FPD6 interaction \cite{fpd6}, where $^{40}$Ca is taken as an inertia
core, and the valence neutrons are distributed in $4$ single-neutron
levels $0f_{7/2}$, $1p_{3/2}$, $0f_{5/2}$, and $1p_{1/2}$.

We first consider the nucleus $^{48}$Ca, where the BCS results in
only a trivial zero solution due to the ``complete filling'' of the
$0f_{7/2}$ orbit. In the Hamiltonian (\ref{H_f}), we keep only the
pairing matrix elements $\langle jj;0|V|j'j';0 \rangle$ of FPD6 for
the two-body interaction $V$, and the single-particle energies
$\epsilon$ are fixed by experimental data as following. From the
spectrum of $^{49}$Ca we read $\epsilon_{p_{1/2}} -
\epsilon_{p_{3/2}} = 2.023$MeV, $\epsilon_{f_{5/2}} -
\epsilon_{p_{3/2}} = 3.585$MeV. And the neutron absorption energy of
$^{48}$Ca gives $\epsilon_{p_{3/2}} = - 5.146$MeV.
$\epsilon_{f_{7/2}}$ is estimated within the single-$j$ degenerate
pairing model as $\epsilon_{f_{7/2}} = -9.945 {\rm MeV} + 0.541 {\rm
MeV} = -9.404 {\rm MeV}$, where $-9.945$MeV is the neutron emission
energy of $^{48}$Ca and $0.541$MeV is the FPD6 pairing strength for
the $0f_{7/2}$ orbit.

\begin{figure*}
\includegraphics[width = 0.8\textwidth]{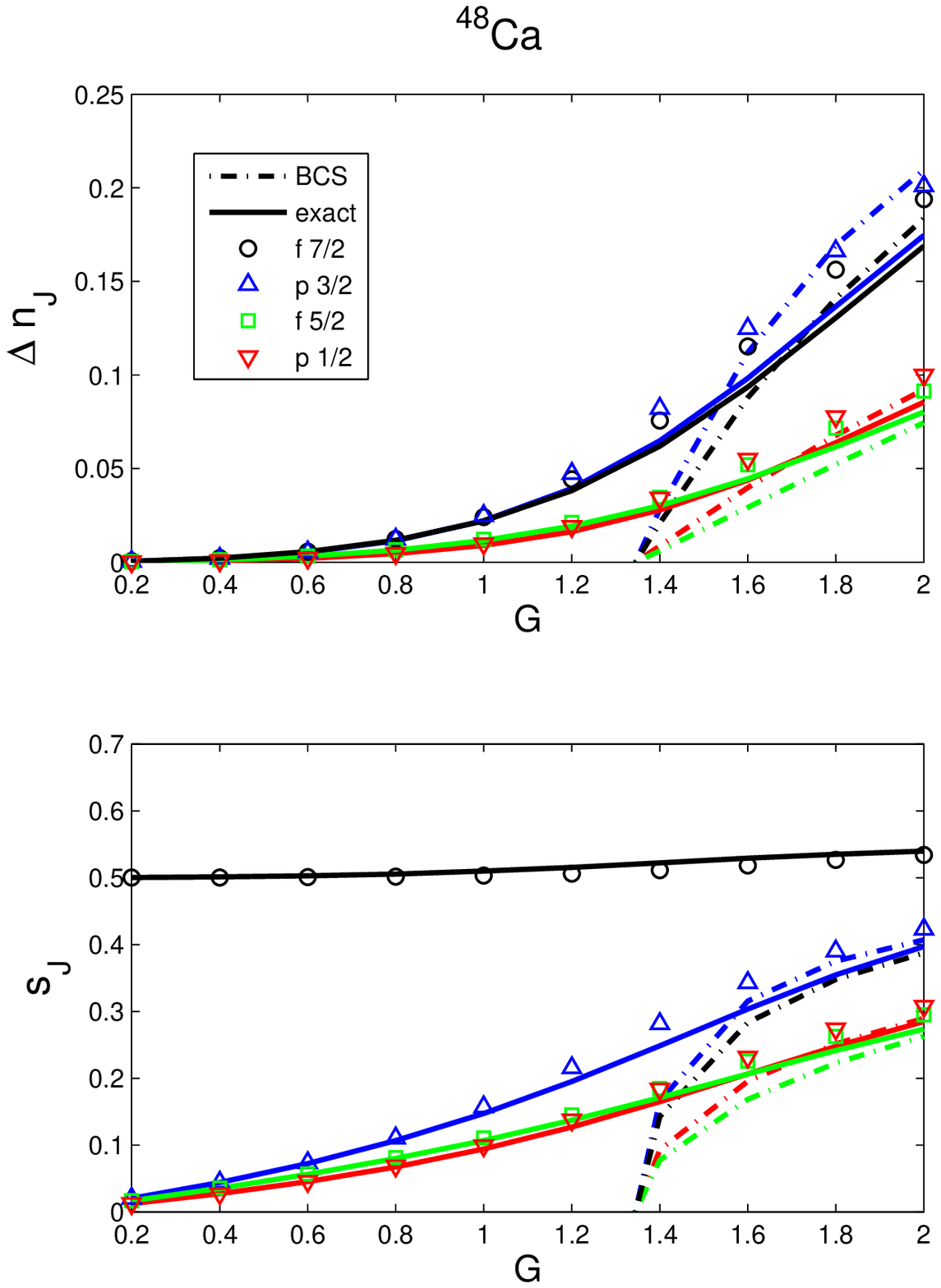}
\caption{\label{Fig_Ca48} (Color online) Occupation numbers $n_J$
(\ref{rho_diag}) and pair emission amplitudes $s_J$
(\ref{kappa_diag}) in $^{48}$Ca as a function of pairing strength
$G$. In the calculation the FPD6 pairing two-body matrix elements
are multiplied by $G$ ($G = 1$ is realistic). The upper panel plots
$\Delta n_J$, the derivations from the naive Fermi occupation
($\Delta n_J = 1-n_{f7/2}, n_{p3/2}, n_{f5/2}, n_{p1/2}$). The solid
lines and dashed-dotted lines show the exact results (by NuShellX)
and BCS results respectively. The symbols show the GDM results,
where black circles, blue up-triangles, green squares, and red
down-triangles are for single-particle levels $f_{7/2}$, $p_{3/2}$,
$f_{5/2}$, and $p_{1/2}$, respectively. The same color convention is
used in plotting the solid lines (exact) and dashed-dotted lines
(BCS). The plotted BCS $s_J$ is defined as $s_1 =~
_{\rm{BCS}}\langle {\phi_N} | a_{\tilde{1}} a_1 | {\phi_N}
\rangle_{\rm{BCS}} = \sqrt{n_1(1-n_1)}$. }
\end{figure*}

The results are given in Fig. \ref{Fig_Ca48}. The realistic case
corresponds to $G = 1$ in the horizontal axis. We see that the GDM
calculation reproduces quite well the exact results (by the
shell-model code NuShellX \cite{MSU_Nu}) of occupation numbers $n_J$
and pair emission amplitudes $s_J$, while BCS fails giving only
trivial zero results. To see how the theory behaves at different
pairing strength, an artificial factor $G$ is introduced that is
multiplied onto the FPD6 pairing two-body matrix elements. We do a
set of calculations at different values of $G$ (from $0.2$ to
$2.0$). The GDM theory does quite well at all pairing strength,
including those below the critical value ($G_c = 1.345$) of BCS. It
even gets one detail right: the inversion (around $G = 1.6$) of
relative positions of the two very close curves for $f_{5/2}$ and
$p_{1/2}$. Because some numbers in Fig. \ref{Fig_Ca48} are very
close and difficult to see, we also list them in Table
\ref{tab_Ca48}.

\begin{figure*}
\includegraphics[width = 0.8\textwidth]{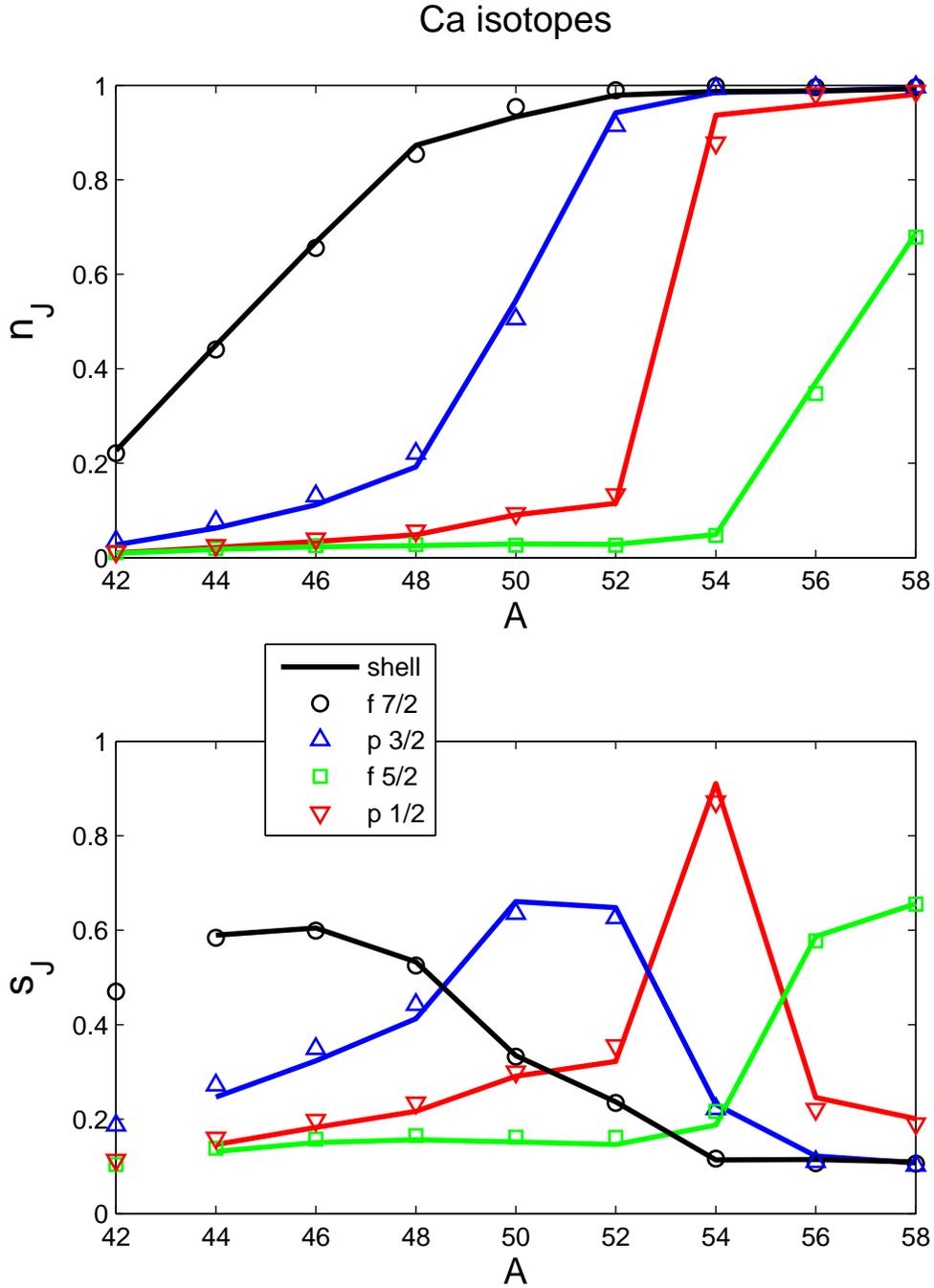}
\caption{\label{Fig_Ca_iso} (Color online) Occupation numbers $n_J$
(\ref{rho_diag}) and pair emission amplitudes $s_J$
(\ref{kappa_diag}) in calcium isotopes ($A$ is the mass number). The
solid lines show the shell-model results (by NuShellX), using the
FPD6 single-particle energies and pairing two-body matrix elements.
The symbols show the GDM results, where black circles, blue
up-triangles, green squares, and red down-triangles are for
single-particle levels $f_{7/2}$, $p_{3/2}$, $f_{5/2}$, and
$p_{1/2}$, respectively. The same color convention is used in
plotting the solid lines (shell). }
\end{figure*}

Next we test the theory in different nuclei. The chain of calcium
isotopes is calculated with mass number $42 \le A \le 58$. For
simplicity in this example we fix the single-particle energies
$\epsilon$ by the FPD6 ones: $\epsilon_{f_{7/2}} = -8.3876$MeV,
$\epsilon_{p_{3/2}} = -6.4952$MeV, $\epsilon_{f_{5/2}} =
-1.8966$MeV, and $\epsilon_{p_{1/2}} = -4.4783$MeV. And in the
two-body interaction $V$ we still keep only the FPD6 pairing matrix
elements. The results are shown in Fig. \ref{Fig_Ca_iso}. The GDM
method reproduce the exact results quite well, even the sudden
changes around $A = 54$.

\section{Summary     \label{Sec_sum}}

In summary, we explored the possibility of using the pair condensate
(\ref{gs}) instead of the quasi-particle vacuum as the starting
point of the GDM method. As the lowest order result, a theory for
nuclear pairing is proposed that conserves the exact particle number
and is valid at arbitrary pairing strength (including those below
the critical point of BCS). Correlations beyond the mean field could
be studied solving higher-order equations in the GDM formalism.

Odd-mass nuclei could be calculated consistently. The effective
Hamiltonian, $\langle 2N+1 | H | 2N + 1\rangle = \langle 2N | a H
a^\dagger | 2N \rangle$, was calculated by substituting Eq.
(\ref{H_f}) into the above expression and then using factorizations
similar to Eq. (\ref{fac1}), where the density matrices $a^\dagger
a$ and $aa$ are known from the neighboring even-even nuclei.
Spectroscopic factors, $\langle 2N-1 | a | 2N \rangle = \langle 2N |
a^\dagger a | 2N \rangle$, could also be
calculated in a similar way. These will be studied in the future.\\

The author gratefully acknowledges discussions with Prof. Vladimir
Zelevinsky.

\newpage
\phantom{a}
\newpage

%

\begin{table*}
  \centering
  \caption{Results of the shell-model and GDM calculations plotted in Fig. \ref{Fig_Ca48}. $n_J$ are occupation numbers, and $s_J$ are pair emission amplitudes.}  \label{tab_Ca48}
  \begin{tabular}{|c|c|c|c|c|c|c|c|c|c|c|}
    \hline
    $G = $             & $0.2$ & $0.4$ & $0.6$ & $0.8$ & $1.0$ & $1.2$ & $1.4$ & $1.6$ & $1.8$ & $2.0$  \\    \hline \hline
    $n_{f_{7/2}}~{\rm{exact}}$ & 0.9996 & 0.9979 & 0.9945 & 0.9883 & 0.9780 & 0.9618 & 0.9381 & 0.9065 & 0.8695 & 0.8311   \\
    ~~~~~~~{\rm{GDM}} & 0.9995  &  0.9979  &  0.9944  &  0.9877  &  0.9759   & 0.9557  &  0.9244  &  0.8848  &  0.8438  &    0.8061  \\ \hline
    $n_{p_{3/2}}~{\rm{exact}}$ & 0.0004 & 0.0019 & 0.0053 & 0.0115 & 0.0222 & 0.0394 & 0.0648 & 0.0983 & 0.1366 & 0.1747   \\
    ~~~~~~~{\rm{GDM}} & 0.0004  &  0.0020  &  0.0054  &  0.0122 &   0.0249  &  0.0474  &  0.0822  &  0.1249  &  0.1663 &    0.2010  \\ \hline
    $n_{f_{5/2}}~{\rm{exact}}$ & 0.0003 & 0.0013 & 0.0032 & 0.0064 & 0.0116 & 0.0192 & 0.0302 & 0.0445 & 0.0616 & 0.0802   \\
    ~~~~~~~{\rm{GDM}} & 0.0003  &  0.0013  &  0.0032  &  0.0066  &  0.0122  & 0.0212  &  0.0346  &  0.0520  &  0.0716  &    0.0913  \\ \hline
    $n_{p_{1/2}}~{\rm{exact}}$ & 0.0001 & 0.0008 & 0.0021 & 0.0045 & 0.0089 & 0.0163 & 0.0278 & 0.0439 & 0.0638 & 0.0856   \\
    ~~~~~~~{\rm{GDM}} & 0.0002  &  0.0007  &  0.0021  &  0.0047  &  0.0098  &  0.0190  &  0.0342  &  0.0549  &  0.0777  &    0.0999 \\    \hline \hline
    $s_{f_{7/2}}~{\rm{exact}}$ & 0.5002 &  0.5010 & 0.5026 & 0.5054 & 0.5097 & 0.5153 & 0.5221 & 0.5291 & 0.5350 & 0.5398    \\
    ~~~~~~~{\rm{GDM}}          & 0.5001 &  0.5003 & 0.5007 & 0.5016 & 0.5032 & 0.5061 & 0.5112 & 0.5183 & 0.5264 & 0.5340    \\ \hline
    $s_{p_{3/2}}~{\rm{exact}}$ & 0.0201 &  0.0438 & 0.0723 & 0.1064 & 0.1472 & 0.1951 & 0.2485 & 0.3035 & 0.3544 & 0.3969    \\
    ~~~~~~~{\rm{GDM}}          & 0.0201 &  0.0441 & 0.0735 & 0.1103 & 0.1571 & 0.2155 & 0.2815 & 0.3430 & 0.3902 & 0.4231    \\ \hline
    $s_{f_{5/2}}~{\rm{exact}}$ & 0.0167 &  0.0353 & 0.0562 & 0.0798 & 0.1067 & 0.1372 & 0.1709 & 0.2064 & 0.2413 & 0.2734    \\
    ~~~~~~~{\rm{GDM}}          & 0.0167 &  0.0354 & 0.0566 & 0.0812 & 0.1103 & 0.1449 & 0.1845 & 0.2253 & 0.2627 & 0.2948    \\ \hline
    $s_{p_{1/2}}~{\rm{exact}}$ & 0.0123 &  0.0270 & 0.0450 & 0.0670 & 0.0940 & 0.1268 & 0.1649 & 0.2061 & 0.2473 & 0.2848    \\
    ~~~~~~~{\rm{GDM}}          & 0.0123 &  0.0272 & 0.0455 & 0.0688 & 0.0988 & 0.1373 & 0.1836 & 0.2313 & 0.2734 & 0.3076    \\ \hline
  \end{tabular}
\end{table*}

%
%
%
%


\begin{thebibliography}{50}

\bibitem{BCS} J. Bardeen, L. N. Cooper, and J. R. Schrieffer, Phys.
Rev. {\bf 106}, 162 (1957); Phys. Rev. {\bf 108}, 1175 (1957).

\bibitem{BCS_nucl1} A. Bohr, B. R. Mottelson, and D. Pines, Phys. Rev.
{\bf 110}, 936 (1958).

\bibitem{BCS_nucl2} S. T. Belyaev, K. Dan. Vidensk. Selsk. Mat. Fys.
Medd. {\bf 31}, (11) (1959).


\bibitem{BCS_book}  Ricardo A Broglia, and Vladimir Zelevinsky, {\it Fifty Years of Nuclear BCS: Pairing in Finite Systems}
(World Scientific, 2013).


\bibitem{kerman63} A. Kerman and A. Klein, Phys. Rev. {\bf 132}, 1326 (1963).

\bibitem{BZ} S.T. Belyaev and V.G. Zelevinsky, Yad. Fiz.
{\bf 11}, 741 (1970) [Sov. J. Nucl. Phys. {\bf 11}, 416 (1970)];
Yad. Fiz. {\bf 16}, 1195 (1972) [Sov. J. Nucl. Phys. {\bf 16}, 657
(1973)]; Yad. Fiz. {\bf 17}, 525 (1973) [Sov. J. Nucl. Phys. {\bf
17}, 269 (1973)].

\bibitem{Zele_ptps} V.G. Zelevinsky, Prog. Theor. Phys. Suppl. {\bf 74-75}, 251 (1983).

\bibitem{shtokman75} M.I. Shtokman, Yad. Fiz. {\bf 22}, 479 (1975) [Sov. J. Nucl. Phys. {\bf 22},
247 (1976)].




\bibitem{Jia_1} L. Y. Jia, Phys. Rev. C {\bf 84}, 024318 (2011).


\bibitem{Jia_2}  L. Y. Jia, and V. G. Zelevinsky , Phys. Rev. C {\bf 84}, 064311
(2011).


\bibitem{Jia_3}  L. Y. Jia, and V. G. Zelevinsky , Phys. Rev. C {\bf 86}, 014315 (2012).



\bibitem{Zele_pairing}  Alexander Volya, and Vladimir Zelevinsky, nucl-th/9912068;
MSUCL-1144 (1999).

\bibitem{fpd6}  W.A. Richter, M.G. Van Der Merwe, R.E. Julies, and B.A.
Brown, Nucl. Phys. {\bf A523}, 325 (1991).

\bibitem{MSU_Nu} NuShellX@MSU, B. A. Brown and W. D. M. Rae,\\
http://www.nscl.msu.edu/~brown/resources/resources.html



%
%
%
%
%
%
%
%
%
%
%
%
%
%
%
%
%
%
%


\end{thebibliography}
\end{document}